\documentclass{elsart1p}

\usepackage{graphicx}

\usepackage{amssymb}

\newcommand{\beq}{\begin{equation}}
\newcommand{\eeq}{\end{equation}}
\newcommand{\bqa}{\begin{eqnarray}}
\newcommand{\eqa}{\end{eqnarray}}
\newcommand{\be}{\begin{equation}}
\newcommand{\ee}{\end{equation}}
\newcommand{\bea}{\begin{eqnarray}}
\newcommand{\eea}{\end{eqnarray}}

\newcommand{\0}{\over }

\newcommand{\6}{\partial }

\begin{document}

\begin{frontmatter}



\title{Nonabelian plasma instabilities in Bjorken expansion}


\author{Anton Rebhan}

\address{Institut f\"ur Theoretische Physik, Technische Universit\"at Wien,\\
        Wiedner Hauptstrasse 8-10, A-1040 Vienna, Austria}

\begin{abstract}
Plasma instabilities are parametrically the dominant nonequilibrium dynamics 
of a weakly coupled quark-gluon 
plasma. In recent years
the time evolution of the corresponding collective colour fields
has been studied in stationary anisotropic situations.
Here I report on recent numerical results on the
time evolution of the most unstable modes
in a longitudinally expanding plasma as they grow
from small rapidity fluctuations to amplitudes where
non-Abelian self-interactions become important.
\end{abstract}


\end{frontmatter}


The experimental findings at the Relativistic Heavy Ion Collider
RHIC \cite{Tannenbaum:2006ch} suggest a much faster 
thermalization than can be accounted for by the (weak-coupling)
bottom-up thermalization scenario of Ref.~\cite{Baier:2000sb}.
However, as first pointed out by Arnold et al.\ \cite{Arnold:2003rq},
this scenario did not include the inevitable presence
of plasma instabilities in an anisotropic quark-gluon 
plasma \cite{Mrowczynski:1988dz,Romatschke:2003ms}. In fact,
these are even the most important collective
effects to leading order. Much effort has by now been invested recently in
studying the specific non-Abelian dynamics of plasma instabilities,
in particular in the crucial regime
where the associated colour fields are still too weak to
affect the distribution of hard particles but already strong enough
for essential non-Abelian self-interactions.

Initial numerical studies that concentrated on
the most unstable modes, namely long-wavelength 
coherent colour fields that are constant in transverse directions,
suggested continued exponential growth also in this nonlinear
regime \cite{Rebhan:2004ur}. Subsequent studies 
\cite{Arnold:2005vb,Rebhan:2005re}
with fully 3+1-dimensional colour fluctuations found however 
a late-time transition from exponential to (rather weak) linear growth.
On the other hand,
the more recent simulations of Ref.~\cite{Bodeker:2007fw} 
found a continued exponential
growth of initially small perturbations
in the case of very strong momentum anisotropy, at least for
small initial field configurations, which is similar to the
behaviour of the 1+1-dimensional case.

All the above-mentioned investigations were performed within the
hard-loop effective theory of a collisionless stationary anisotropic
plasma \cite{Mrowczynski:2004kv}. Clearly, in view of the crucial
dependence of the fate of non-Abelian plasma instabilities on the
amount of momentum anisotropy, which rapidly increases 
during the free-streaming
stage, a generalization to Bjorken expansion is desirable.\footnote{%
In the color-glass framework, the issue of instabilities in an
expanding system was
studied in Ref.~\cite{Romatschke:2006nk}.}

In the stationary anisotropic situation, 
the coloured fluctuations of a colour-neutral space-time-independent
background of hard particles can be factored according to
\be
\delta f^a(x;p)=-g W^a_\mu(t,\mathbf x;\mathbf v)
\6_{(p)}^\mu f_0(\mathbf p)
\ee
where the auxiliary fields $W$ depend only on velocities 
$v^\mu\equiv p^\mu/|\mathbf p|=(1,\mathbf v)$
and are governed by a Vlasov equation
\be
[v\cdot D(A)]W_\mu(x;\mathbf v)
= F_{\mu\gamma}(A) v^\gamma \, 
\ee
in which the hard momentum scale $p^0$ does not appear.
The latter can be ``integrated out'', determining just
the mass parameter in the induced current $j^\mu[\delta f]
=m_D^2 \int_{\mathbf v}\mathcal W(x;\mathbf v)$,
where $\mathcal W$ is one particular linear combination of the 
components $W^\nu$.
The collective colour fields associated with plasma instabilities
can then be determined by solving the equation for $\mathcal W$
together with the non-Abelian Maxwell equations
$D_\rho(A)F^{\rho\mu}=j^\mu(x)$.

Remarkably, this formalism can largely be taken over to
the nonstationary case of a plasma undergoing Bjorken expansion
\cite{Romatschke:2006wg}.

In comoving (rapidity) variables
$x^\alpha=(\tau=\sqrt{t^2-z^2},x^i,\eta={\rm atanh}(z/t))$
and $p^\alpha=|\mathbf p_\perp|
(\cosh(y-\eta),\cos\phi, \sin\phi,\tau^{-1}{\sinh(y-\eta)}$
with $y={\rm atanh}(p^0/p^z)$,
a free-streaming background distribution is given by
\be\label{faniso}
f_0(\1p,x)=f_{\rm iso}\left(\sqrt{p_\perp^2+p_\eta^2/\tau_{\rm iso}^2}\right)
=f_{\rm iso}\left(\sqrt{p_\perp^2+(p'^z\tau/\tau_{\rm iso})^2}\right)
\ee
where $p'_z$ is the boosted longitudinal momentum.

One can again introduce auxiliary fields $W$
\be
\delta f^a(x;p)=-g W^a_\alpha(\tau,x^{i},\eta;\phi,y)
\6_{(p)}^\alpha f_0(p_\perp,p_\eta)
\ee
that obey 
$
v\cdot D\, W_\alpha(\tau,x^{i},\eta;\phi,y)|_{\phi,y}=v^\beta F_{\alpha\beta},
$
but now
\be
v^\alpha\equiv {p^\alpha / |{\mathbf p_\perp}|}
=(\cosh(y-\eta),\cos\phi, \sin\phi,\tau^{-1}{\sinh(y-\eta)}).
\ee

The induced current in the non-Abelian Maxwell equations 
is given by
\be
\frac1\tau D_\alpha(\tau F^{\alpha\beta})=j^\beta(\tau,x^i,\eta)=
-{m_D^2(\tau\!=\!\tau_{\rm iso})\02}\int_0^{2\pi} {d\phi \0 2\pi}
\int_{-\infty}^\infty dy \,
v^\beta
\mathcal W(\tau,x^i,\eta;\phi,y)
\ee
with $\mathcal W=v^i W_i-{\tau\0\tau_{\rm iso}^2} \sinh(y-\eta)\,  W_\eta$.

In numerical simulations one now ends up discretizing velocity/rapidity
space that has the form of a cylinder parametrized by $\phi,y$ 
instead of a sphere ${\bf v}^2=1$.
These numerical simulations have to start at a finite proper time $\tau=\tau_0$, both for practical reasons and for the physical reason that a plasma
description makes sense at the earliest around $\tau\sim Q_s^{-1}$, where
$Q_s$ is the saturation scale of perturbative QCD. 
Having chosen a background distribution function of the form
(\ref{faniso}), the only free parameters are $\tau_{\rm iso}$, the
(possibly fictitious) time when $f_0$ is momentarily isotropic, and
the Debye mass following from $f_0$ at that point in time.

In the results displayed in Fig.~1, $\tau_{\rm iso}=0.1\tau$, i.e.\
the distribution starts out already in oblate form
and becomes increasingly oblate as time goes on. The mass parameter
is matched to results from the saturation (colour-glass) scenario
as explained in the appendix. In this numerical calculation, only
the (most unstable) modes which are transversely constant and thus
effectively 1+1-dimensional have been considered. These instabilities
have been seeded by small rapidity fluctuations 
with a spectrum modelled after Ref.~\cite{Fukushima:2006ax}.
Given the findings in stationary anisotropic plasmas, the
growth of non-Abelian plasma instabilities found here probably
gives an upper bound on more generic cases. Considering that
for $Q_s\simeq 1$ GeV for RHIC and $\simeq 3$ GeV for the LHC and that thus
the maximal time in Fig.~1 corresponds to some 20 fm/c for RHIC
and 7 fm/c for the LHC, one finds an uncomfortably long delay for the
onset of plasma instabilities at least for RHIC. Further studies
of more generic initial conditions (including then also strong
initial fields) is work in progress.

\begin{figure}
\centerline
{\includegraphics[
clip,
width=0.64\linewidth]{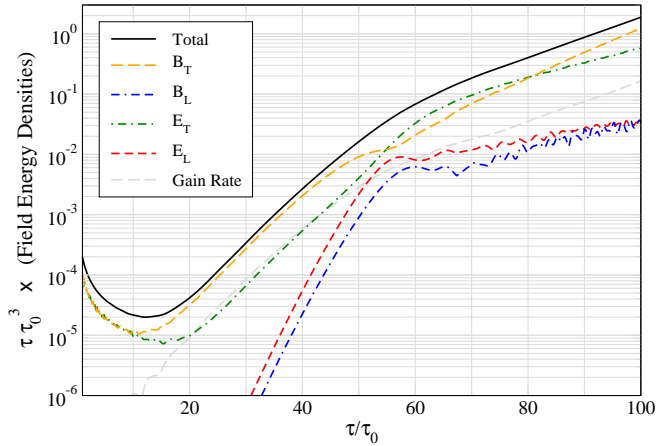}~~}
\caption{\label{fig7}
Results \cite{Rebhan:2008uj}
from a 1D+3V real-time lattice simulation of
non-Abelian plasma instabilities in Bjorken expansion, seeded by small initial
rapidity fluctuations with a spectrum modelled after Ref.~\cite{Fukushima:2006ax}.  
We show the proper-time dependence
of the total chromo-field energy density and its individual components
$\mathcal E=\mathcal E_{B_T}+\mathcal E_{E_T}+\mathcal E_{B_L}+\mathcal E_{E_L}
=\mathcal E_T+\mathcal E_L$
as well as the chromo-field energy gain rate $R$ defined by
$R={d{\cal E}}/{d\tau} + {2}{\cal E}_T/{\tau}.$
}
\end{figure}

\appendix
\section{Parameters of the numerical simulation}

For fixing the dimensionful parameters of the numerical
simulation in a way
that makes contact with heavy-ion physics, the
so-called Colour-Glass-Condensate framework
\cite{McLerran:1993ni,Iancu:2003xm} is invoked, and as starting
time for the plasma phase $\tau_0\simeq
Q_s^{-1}$ is chosen, where $Q_s$ is the so-called saturation scale.

The only other dimensionful parameter, the Debye mass
$m_D$ at the (fictitious because pre-plasma) time $\tau_{\rm iso}$,
is fixed by
assuming a squashed Bose-Einstein distribution function for the
hard particle distribution function (\ref{faniso}) with
$f_{\rm iso}(p)=\mathcal N (2N_g) / (e^{p/T}-1)$ where $N_g=N_c^2-1$ 
is the number
of gluons and $\mathcal N$ a normalization
that is adjusted such that at $\tau=\tau_0$ the hard-gluon density
of CGC estimates is matched. 
Since the expansion is by assumption purely longitudinal, $T$ is a constant
transverse temperature, and it has indeed been found in
CGC calculations that the gluon distribution is approximately thermal
in the transverse directions, with $T=Q_s/d$ and $d^{-1}\simeq 0.47$
according to Ref.~\cite{Iancu:2003xm}.
The normalization $\mathcal N$ can then be fixed by
following Ref.~\cite{Baier:2002bt}, who
write the initial hard-gluon density as
\be
n(\tau_0)=c \frac{N_g Q_s^3}{4\pi^2 N_c\alpha_s (Q_s \tau_0)},
\ee 
where $c$ is
the gluon liberation factor, for which different estimates can be extracted
from the literature. 

We adopted the value $c=2\ln2\approx 1.386$
from an analytical estimate in Ref.~\cite{Kovchegov:2000hz}
which turned out to be fairly close to
the most recent numerical result $c\simeq 1.1$
in Ref.~\cite{Lappi:2007ku}. 
With $\tau_{\rm iso}$ remaining a free parameter which determines how
anisotropic the gluon distribution is at $\tau_0$, 
the normalization $\mathcal N$ is now fixed by 
$
n(\tau_0)\frac{\tau_0}{\tau_{\rm iso}}=
n(\tau_{\rm iso})=\frac{2 \zeta(3)}{\pi^2}\mathcal N N_g T^3.
$
For a purely gluonic plasma, the isotropic Debye mass is given by
$
m_D^2(\tau_{\rm iso})=\mathcal N {4\pi \alpha_s N_c T^2}/{3},
$
which together leads to
\be
m_D^2(\tau_{\rm iso})\tau_0^2(Q_s \tau_0)^{-1}= \frac{\pi c d }{6 \zeta(3)}
\frac{\tau_0}{\tau_{\rm iso}} 
\approx 1.285
\frac{\tau_0}{\tau_{\rm iso}},\;\;
\mathcal N\approx 0.4628\, \alpha_s^{-1} (Q_s \tau_0)^{-1} \frac{\tau_0}{\tau_{\rm iso}}, 
\ee
when $c=2\ln 2$ and $N_c=3$. We adopt this value for our simulations
where $N_c=2$, since in previous studies of the stationary anisotropic
situation little difference was found between the SU(2) and the SU(3) case
provided $m_D$ was the same \cite{Rebhan:2005re}.
With our choice of an initial anisotropy given by $\tau_0/\tau_{\rm iso}=10$,
equating $\tau_0=Q_s^{-1}$ and using units where $\tau_0=1$,
the above result corresponds to the value $m_D=3.585$ employed in
Fig.~\ref{fig7}. This corresponds to
an energy density of hard particles $\mathcal E(\tau_{\rm iso}\!=0.1\tau_0)=
\mathcal N 8\pi^2 Q_s^4 d^{-4}/15
$
and (using Eq.~(15) of \cite{Rebhan:2008uj}) $\mathcal E(\tau_0)\approx
0.0789\, \mathcal E(\tau_{\rm iso})
\approx 0.0938\, Q_s^4(Q_s\tau_0)^{-1}\alpha_s^{-1}$.








\end{document}